\documentclass[aps,prc,twocolumn,showpacs,nofootinbib,aps,letterpaper]{revtex4-1}

\usepackage{amsmath}
\usepackage{amsfonts,amssymb}
\usepackage{multirow}
\usepackage[dvips]{graphicx}
\usepackage{float}
\usepackage{appendix}
\usepackage{color}
\usepackage{url}
\usepackage{subfigure}

\def\beq{\begin{equation}}
\def\eeq{\end{equation}}
\def\bea{\begin{eqnarray}}
\def\eea{\end{eqnarray}}

\def\nl{\nonumber\\}
\def\jpsi{J/\psi}
\def\epp{\eta'\pi\pi}
\def\eppc{\eta'\pi^+\pi^-}

\def\ppbar{\bar p p}
\def\nnbar{\bar n n}
\def\NNbar{\bar N N}

\RequirePackage{xspace} \allowdisplaybreaks

\begin{document}

\title{$J/\psi \to \gamma\eta'\pi^+\pi^-$ and the structure observed around the $\bar pp$ threshold}
\author{Ling-Yun Dai$^{1,2}$}      
\author{Johann Haidenbauer$^2$}    
\author{Ulf-G. Mei{\ss}ner$^{3,2}$}
\affiliation{$^1$School of Physics and Electronics, Hunan University, Changsha 410082, China\\
$^2$Institute for Advanced Simulation, J\"ulich
Center for Hadron Physics, and Institut f\"ur
Kernphysik, Forschungszentrum J\"ulich, D-52425 J\"ulich, Germany\\
$^3$Helmholtz-Institut f\"ur Strahlen- und Kernphysik and Bethe
Center for Theoretical Physics, Universit\"at Bonn, D-53115 Bonn,
Germany}

\begin{abstract}
We analyze the origin of the structure observed in the reaction $J/\psi \to \gamma \eta'\pi^+\pi^-$
for $\eta'\pi^+\pi^-$ invariant masses close to the antiproton-proton ($\bar pp$) threshold,
commonly associated with the $X(1835)$ resonance. 
Specifically, we explore the effect of a possible contribution from the two-step process 
$J/\psi \to \gamma \bar NN \to \gamma \eta'\pi^+\pi^-$. 
The calculation is performed in distorted-wave Born approximation which allows an appropriate
inclusion of the $\bar NN$ interaction in the transition amplitude. The $\bar NN$ amplitude
itself is generated from a corresponding potential recently derived within chiral effective 
field theory. 
We are able to reproduce the measured spectra for the reactions $J/\psi \to \gamma \bar pp$ 
and $J/\psi \to \gamma \eta'\pi^+\pi^-$ for invariant masses around the $\bar pp$ threshold. 
The structure seen in the $\eta'\pi^+\pi^-$ spectrum emerges as a threshold effect due to the 
opening of the $\bar pp$ channel. 
\end{abstract}


\maketitle

\section{Introduction}
\label{sec:introduction}

The $X(1835)$ resonance, first discovered by the BES Collaboration in 2005
in the decay $\jpsi \to \gamma\eppc$ \cite{Ablikim:2005} and subsequently
seen in other reactions \cite{Ablikim:2013,Ablikim:2015,Ablikim:2018},
but only faintly by other groups \cite{Zhang:2012,He:2013}, has a long and winding history.
Initially the resonance was associated with the anomalous near-threshold
enhancement in the antiproton-proton ($\ppbar$) invariant mass spectum
in the reaction $\jpsi \to \gamma\ppbar$ \cite{Bai:2003,Ablikim:2012}
which would point to a baryonium-type state (or $\NNbar$ quasi-bound state)
as possible explanation for its structure.
However, with increasing statistics \cite{Ablikim:2010}
it became clear that the two phenomena are not necessarily connected, not least
due to a striking difference in the width of the respective resonances required
for describing the invariant mass spectra of the two reactions in question.
Yet another facet was added in the most recent publication of the
BESIII Collaboration on the decay $\jpsi \to \gamma\eppc$ \cite{Ablikim:2016}.
Now the initial peak around $1835$~MeV is practically gone but has
reappeared as a structure that is located very close to the
$\ppbar$ threshold, namely around $1870$~MeV.

A more detailed coverage of the historical developement regarding the
$X(1835)$ resonance can be found in recent summary papers
\cite{BesRev,Xu:2017}. These works provide also an overview of
the large amount of theoretical investigations performed in the context of
the $X(1835)$. Naturally, in many of them an interpretation of the
resonance in terms of a baryonium state is the key element. Indeed, some of
these studies attempt to establish a
direct and quantitative connection of the resonance with predictions
of $\NNbar$ potentials that were fitted to $\ppbar$ scattering data
\cite{Dedonder:2009,Milstein:2017}.

In the present work we aim at a quantitative analysis of the most recent
BESIII data on the reaction $\jpsi \to \gamma\eppc$ \cite{Ablikim:2016}.
The study is based on the hypothesis that the structure seen in the invariant
mass spectrum is indeed linked with the opening of the $\ppbar$ channel.
The incentive for that comes from past studies of $e^+e^-$ annihilation
into multipion states. Also in this case, and specifically in the reactions
$e^+e^-\to 3(\pi^+\pi^-)$, $2(\pi^+\pi^-\pi^0)$, $\omega\pi^+\pi^-\pi^0$,
and $e^+e^-\to 2(\pi^+\pi^-)\pi^0$, structures were observed in the
experiments at energies around the $\ppbar$ threshold \cite{BABAR,BABAR2,CMD,CMD2}.
Calculations by our group \cite{Haidenbauer:2015} and others \cite{Milstein:2015} suggested
that two-step processes $e^+e^-\to \bar NN \to$ multipions could play an important role
and their inclusion even allowed one to reproduce the data quantitatively near
the $\bar NN$ threshold. Accordingly, the structures seen in the experiments found
a natural explanation as a threshold effect due to the opening of the $\bar NN$ channel,
for the majority of the measured channels.

As already indicated above, with the new $\jpsi \to \gamma\eppc$ data \cite{Ablikim:2016}
the region of interest is now shifted
likewise to energies around the $\ppbar$ threshold. Accordingly, we investigate
the significance of the $\bar NN$ channel for the reaction $\jpsi \to \gamma\eppc$.
Since the decay $\jpsi \to \gamma\ppbar$ constitutes one segment of the assumed
two-step process (the other being $\ppbar \to \eppc$), we reconsider this decay
process in the present paper. Indeed, we had already shown in earlier studies that
it is possible to describe the large near-threshold enhancement observed in the
reaction $\jpsi \to \gamma\ppbar$ by the final-state interaction (FSI)
provided by the $\NNbar$ interaction \cite{Sibirtsev:2004,Haidenbauer:2006,Kang:2015},
see also Refs.~\cite{Dedonder:2009,Milstein:2017,Entem,Chen,Dedonder:2018}.

A main ingredient of our present calculation is the $\NNbar$ interaction.
Here we build on our latest $\NNbar$ potential, derived in the framework
of chiral effective field theory (EFT) up to next-to-next-to-next-to-leading order
(N$^3$LO) \cite{Dai:2017}. That potential reproduces the amplitudes determined in a
partial-wave analysis (PWA) of $\ppbar$ scattering data \cite{Zhou} from the $\NNbar$
threshold up to laboratory energies of $T_{lab} \approx 200-250$~MeV \cite{Dai:2017}.

The paper is structured in the following way:
In Sect.~\ref{sec:formalism} an overview of the employed formalism is provided. 
Sect.~\ref{sec:ppbar} is devoted to the reaction $\jpsi \to \gamma\ppbar$, the
first segment of the considered two-step process.  
In particular, a comparison with the $\jpsi \to \gamma\ppbar$ data from the 
BESIII Collaboration is presented. As in our initial study \cite{Kang:2015},   
a refit of the $\NNbar$ amplitudes in the $^1S_0$ partial-wave with 
isospin $I=1$ is required. 
The second segment of the considered two-step process, the reaction $\ppbar \to \eppc$, 
is discussed in Sect.~\ref{sec:results}. 
However, the main focus of this section is on the reaction $\jpsi \to \gamma\eppc$ 
and results for the $\eppc$ invariant mass spectrum are presented. It turns out
that the structure observed in the BESIII experiment at invariant masses near the 
$\NNbar$ threshold is very well reproduced, once effects due to the coupling to 
the $\NNbar$ channel are explicitly taken into account. In view of that observation,
and in the light of the conjectured $X(1835)$ resonance, 
the employed $\NNbar$ interactions are examined with regard to possible bound states. 
The paper ends with concluding remarks. 

\section{Formalism}
\label{sec:formalism}

Our study of the processes $\jpsi \to \gamma \ppbar$ and $\jpsi \to \gamma \eppc$
is based on the distorted-wave Born approximation (DWBA).
It amounts to solving the following set of formally coupled equations:
\begin{eqnarray}
\nonumber
T_{\bar NN\to \bar NN} &=&  V_{\bar NN\to \bar NN} + V_{\bar NN\to \bar NN} G_0 T_{\bar NN\to \bar NN}, \\
\nonumber
T_{\bar NN\to  \epp  } &=&  V_{\bar NN\to  \epp    } + T_{\bar NN\to \bar NN} G_0 V_{\bar NN\to  \epp  }, \\
\nonumber
A_{\jpsi \to \gamma\bar NN} &=&  A^0_{\jpsi \to \gamma\bar NN} + A^0_{\jpsi\to \gamma\bar NN} G_0 T_{\bar NN\to \bar NN}, \\
&& \label{eq:LSA} \\ \nonumber
A_{\jpsi \to  \gamma\epp  } &=&  A^0_{\jpsi \to  \gamma\epp} + A^0_{\jpsi\to \gamma\bar NN} G_0 T_{\bar NN\to \epp  }  \\
\nonumber
                       &=&  A^0_{\jpsi \to \gamma\epp} + A_{\jpsi \to \gamma\bar NN} G_0 V_{\bar NN\to  \epp   }.  \\
&& \label{eq:LS}
\end{eqnarray}
The first line in Eq.~(\ref{eq:LSA}) is the Lippmann-Schwinger equation from which the $\NNbar$
scattering amplitude ($T_{\NNbar}$), is obtained, for a specific $\NNbar$ potential $V_{\NNbar}$,
see Refs.~\cite{NNLO,Dai:2017} for details. The quantity $G_0$ denotes the free $\NNbar$ Green's function.
The second equation defines the amplitude for $\NNbar$ annihilation into the $\eppc$ channel
while the third equation provides the $\jpsi \to \gamma\bar NN$ transition amplitude.
Finally, Eq.~(\ref{eq:LS}) defines the $\jpsi \to \gamma\eppc$ amplitude.
The quantities $A^0_\nu$ denote the elementary (or primary) decay amplitudes for
$\jpsi$ to $\gamma\bar NN$ or $\gamma\epp$.

General selection rules \cite{Kang:2015} but also direct experimental evidence \cite{Ablikim:2015}
suggest that the specific (and unique) $\NNbar$ partial wave that plays a role for energies
around the $\ppbar$ threshold is the $^1S_0$.
For it the equation for the amplitude $A_{\jpsi \to \gamma\bar NN}$ reads \cite{Kang:2015}
\begin{equation}\label{eq:fullFSI}
A = A^0 +\int_0^\infty \frac{dp p^2}{(2\pi)^3} A^0 \frac{1}{2E_k-2E_p+i0^+}T(p,k;E_k) ,
\end{equation}
where $k$ and $E_k$ are the momentum and energy of the proton (or antiproton) in the center-of-mass
system of the $\NNbar$ pair, i.e. $E_k= \sqrt{m_p^2 + k^2}$, with $m_p$  the proton (nucleon)
mass. The subscript of $A$ indicating the channel is omitted in Eq.~(\ref{eq:fullFSI}) for 
simplicity. 
 
The $\NNbar$ $T$-matrix that enters Eq.~(\ref{eq:fullFSI}) fulfils
\begin{eqnarray}
&&T(p',k;E_k)=V(p',k)+ \nl
&& \int_0^\infty \frac{dpp^2}{(2\pi)^3} \, V(p',p)
\frac{1}{2E_k-2E_p+i0^+}T(p,k;E_k)~,\nl
\label{LS}
\end{eqnarray}
where $V$ represents the $\NNbar$ potential in the $^1S_0$ partial wave.

Following the strategy in Ref.~\cite{NNLO,Dai:2017}, the elementary annihilation potential for $\NNbar \to\eppc$
and the transition amplitude $A^0_{\jpsi\to \gamma\bar NN}$ are parameterized by
\begin{eqnarray}
\label{eq:VNN}
V_{\NNbar \to \epp}(q)&=&\tilde C_{\epp} + C_{\epp} q^2, \\
A^0_{\jpsi \to \gamma\bar NN}(q)&=&\tilde C_{\jpsi \to \gamma\bar NN} + C_{\jpsi \to \gamma\bar NN} q^2, 
\label{eq:A0N}
\end{eqnarray}
i.e. by two contact terms analogous to those that arise up to next-to-next-to-leading order (N$^2$LO)
in the treatment of the $\NNbar$ interaction within chiral EFT \cite{Dai:2017}.
The quantity $q$ in Eq.~(\ref{eq:VNN}) is the center-of mass (c.m.) momentum in the $\NNbar$ system.
Note that we multiply the transition potentials in Eqs.~(\ref{eq:VNN}) and (\ref{eq:A0N})
with a regulator (of exponential type) in the actual calculations. This is done consistently with
the $\NNbar$ potentials in Ref.~\cite{Dai:2017} where such a regulator is included.
We also employ the same cutoff parameter as in the $\NNbar$ sector.
Since the threshold for the $\epp$ channel lies significantly below
the one for $\NNbar$, the mesons carry - on average - already fairly high momenta.
Thus, the dependence of the annihilation potential on those momenta should be small
for energies around the $\NNbar$ threshold and it is, therefore, neglected \cite{Kang:2015}.
The constants $\tilde C_\nu$ and $C_\nu$ can be determined by a fit to the $\NNbar \to \epp$
cross section (and/or branching ratio) and the $\jpsi\to \gamma\ppbar$ invariant mass
spectrum, respectively.

The term $A^0_{\jpsi \to\gamma\eppc}$ is likewise parameterized in the form~(\ref{eq:A0N}),
but as a function of the $\epp$ invariant mass $Q$, 
\begin{eqnarray}
A^0_{\jpsi \to \gamma\epp}(Q)&=&\tilde C_{\jpsi \to \gamma\epp} + C_{\jpsi \to \gamma\epp}\,Q.
\label{eq:A0e}
\end{eqnarray}
The arguments for neglecting the dependence on the individual meson momenta are the same as above 
and they are valid again, of course, only for energies around the $\NNbar$ threshold. However, since 
in the $\epp$ case this term represents a background amplitude rather than a transition potential we 
allow the corresponding constants to be complex valued, to be fixed by a fit to 
the $\jpsi \to\gamma\eppc$ event rate.

The explicit form of Eq.~(\ref{eq:LS}) reads
\begin{eqnarray}\label{eq:epp}
&& A_{\gamma\epp, \jpsi}(X;Q)= A^0_{\gamma\epp,\jpsi}(X;Q) + \int_0^\infty \frac{dq q^2}{(2\pi)^3} \nl
&&\qquad \times V_{\epp,\ppbar}(X,q) \frac{1}{Q-2E_q+i0^+}A_{\gamma\ppbar,\jpsi}(q;Q), \nl
\end{eqnarray}
written in matrix notation.
The quantity $X$ stands here symbolically for the momenta in the $\epp$ system. But since we assumed 
that the transition potential does not depend on those momenta, cf. Eqs.~(\ref{eq:VNN}) and (\ref{eq:A0e}), 
$X$ does not enter anywhere into the actual calculation of the amplitudes.
All amplitudes (and the potential) can be written and evaluated as functions of the c.m. momenta 
in the $\NNbar$ ($q$) system and of the invariant mass $Q$ in the $\epp$ system, where the
latter is identical to the energy in the $\NNbar$ subsystem. 

Since the amplitudes do not depend on $X$ the integration over the three-meson phase space can
be done separately when the cross section or the invariant mass spectrum  are calculated.
In practice, it amounts only to a multiplicative factor and, moreover, to a factor that is
the same for the $\NNbar \to \epp$ cross section and the $\jpsi \to \gamma\epp$ invariant mass
spectrum for a fixed value of $Q$. We perform this phase space integration numerically.

Of course, ignoring the dependence of $A^0_{\jpsi\to\gamma\epp}$ on the $\epp$ momenta is only
meaningful for energies around the $\NNbar$ threshold. We cannot extend our calculation down
to the threshold of the $\epp$ channel. However, one has to keep in mind that also the
validity of our $\NNbar$ interaction is limited to energies not too far away from the $\NNbar$
threshold.

The differential decay rate for the processes $\jpsi \to \gamma\ppbar$ can be written in the 
form \cite{Kang:2015,PDG}
\beq \label{eq:decayrate}
\frac{d\Gamma}{dQ}=\frac{\lambda^{1/2}(m_{\psi}^2,Q^2,m_x^2)\sqrt{Q^2-4m_p^2}}{2^7\pi^3m_{\psi}^3}|
\mathcal{M}_{\jpsi\to\gamma\ppbar}|^2 \ ,
\eeq
after integrating over the angles.
Here the K\"all\'en function $\lambda$ is defined as $\lambda(x,y,z)=(x-y-z)^2-4yz$,
$Q\equiv M_{\ppbar}$ is the invariant mass of the $\ppbar$ system, $m_{\psi}$, $m_p$, $m_x$ are the
masses of the $\jpsi$, the proton, and the meson (or photon) in the final state, in order,
while $\mathcal{M}$ is the total Lorentz-invariant reaction amplitude.
The relation between the $A$'s in Eq.~(\ref{eq:LSA}) and (\ref{eq:LS})
and the Lorentz-invariant amplitudes $\mathcal{M}$ for the various reactions is \cite{Joachain}:
\bea
\mathcal{M}_{\NNbar\to\NNbar}&=&-8\pi^2 E_N^2 ~T_{\NNbar\to\NNbar}\;,\nonumber\\
\mathcal{M}_{\jpsi\to\gamma\ppbar}&=&-8\pi^2 E_N \sqrt{E_\gamma E_{\jpsi}} ~A_{\jpsi\to\gamma\ppbar}\;,\nonumber\\
\mathcal{M}_{\NNbar\to\eta'\pi^+\pi^-}&=&-32\sqrt{\pi^7} E_N \sqrt{E_{\eta'} E_{\pi^+} E_{\pi^-}}\nonumber\\ 
&&A_{\NNbar\to\eta'\pi^+\pi^-}\;,\nonumber\\
\mathcal{M}_{\jpsi\to\gamma\eta'\pi^+\pi^-}&=&-32\sqrt{\pi^7} \sqrt{E_\gamma E_{\jpsi} E_{\eta'} E_{\pi^+} E_{\pi^-}} \nonumber\\
&&A_{\jpsi\to\gamma\eta'\pi^+\pi^-}\;.
\eea  
The energies in the reactions $\jpsi\to\gamma\ppbar$, $\NNbar\to\eppc$, and $\jpsi\to\gamma\eppc$
are given by 
\bea
E_N&=&Q/2\;,\nonumber\\
E_{\jpsi}&=&\frac{m_{\psi}^2+Q^2}{2{Q}}\;,\nonumber\\
E_{\gamma}&=&\frac{m_{\psi}^2-Q^2}{2{Q}}\;.\nonumber\\
E_{\eta'}&=&\frac{Q^2-t_1+m_{\eta'}^2}{2Q}\;.\nonumber\\
E_{\pi^+}&=&\frac{Q^2-t_2+m_{\pi}^2}{2Q}\;.\nonumber\\
E_{\pi^-}&=&\frac{t_1+t_2-m_{\pi}^2-m_{\eta'}^2}{2Q}\;,\nonumber
\eea
where $Q$ is either the energy in the $\NNbar$ system or
the invariant mass of the $\ppbar$ or ${\eta'\pi^+\pi^-}$ systems
($M_{\ppbar}$ or $M_{\eta'\pi^+\pi^-}$), 
$t_1=M_{\pi^+\pi^-}^2$, and $t_2=M_{\pi^-\eta'}^2$. 
 
In Eq.~\eqref{eq:decayrate} it is assumed that averaging over the spin states 
has been already performed. Anyway, in the present manuscript we will consider 
only individual partial wave amplitudes.
The cross section for the reaction $\ppbar\to\eta'\pi^+\pi^-$ is given by
\bea
\sigma(\ppbar\to\eta'\pi^+\pi^-)&=&\int_{t_1^-}^{t_1^+}d t_1\int_{t_2^-}^{t_2^+}
\frac{d t_2 \ |\mathcal{M}_{\ppbar\to\eta'\pi\pi}|^2}{1024\pi^3Q^3\sqrt{Q^2-4m_p^2}}\;,\nonumber\\
\eea
where 
\bea
t_1^-&=&4m_\pi^2\;,\nonumber\\
t_1^+&=&(Q-m_{\eta'})^2\;, \nonumber\\
t_2^-&=&\frac{1}{4t_1}\left( (Q^2-m_{\eta'}^2)^2-[\lambda^{1/2}(Q^2,t_1,m_{\eta'}^2)\right.\nonumber\\
&&\left.+\lambda^{1/2}(t_1,m_{\pi}^2,m_{\pi}^2)]^2 \right)\;,\nonumber\\
t_2^+&=&\frac{1}{4t_1}\left( (Q^2-m_{\eta'}^2)^2-[\lambda^{1/2}(Q^2,t_1,m_{\eta'}^2)\right.\nonumber\\
&&\left.-\lambda^{1/2}(t_1,m_{\pi}^2,m_{\pi}^2)]^2 \right)\;.\nonumber\\
\eea
The decay rate for $\jpsi \to \gamma\eppc$ is given by
\bea
\frac{d\Gamma}{d Q}&=&\int_{t_1^-}^{t_1^+}d t_1\int_{t_2^-}^{t_2^+}d t_2
\frac{(m_{\psi}^2-Q^2)|\mathcal{M}_{\jpsi \to \gamma\eppc}|^2}{6144\pi^5 m_{\psi}^3 Q}\;.\nonumber\\
\eea

\vskip 0.5cm

\begin{figure}[htbp]
\begin{center}
\includegraphics[height=50mm,clip]{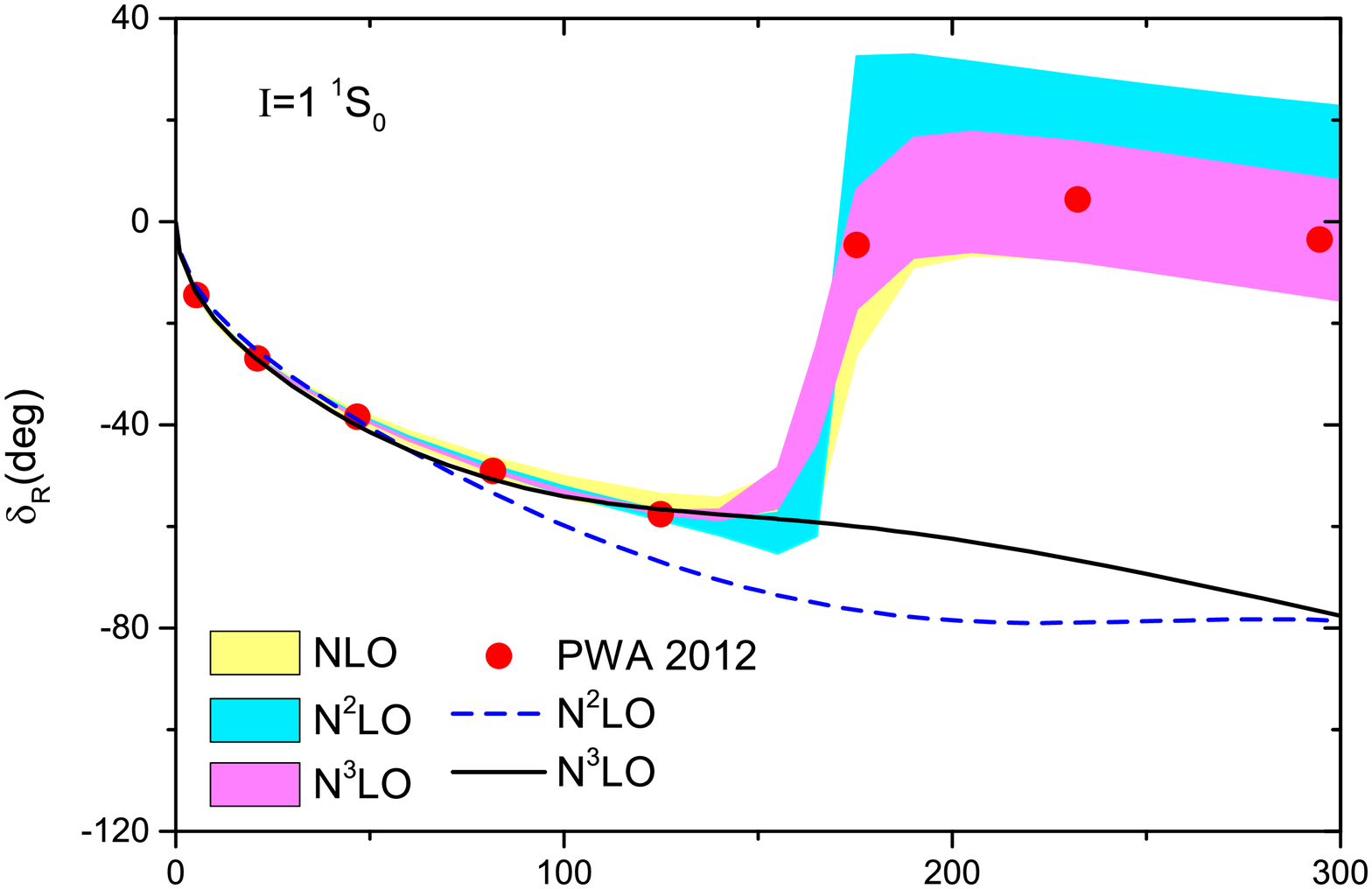}
\includegraphics[height=55mm,clip]{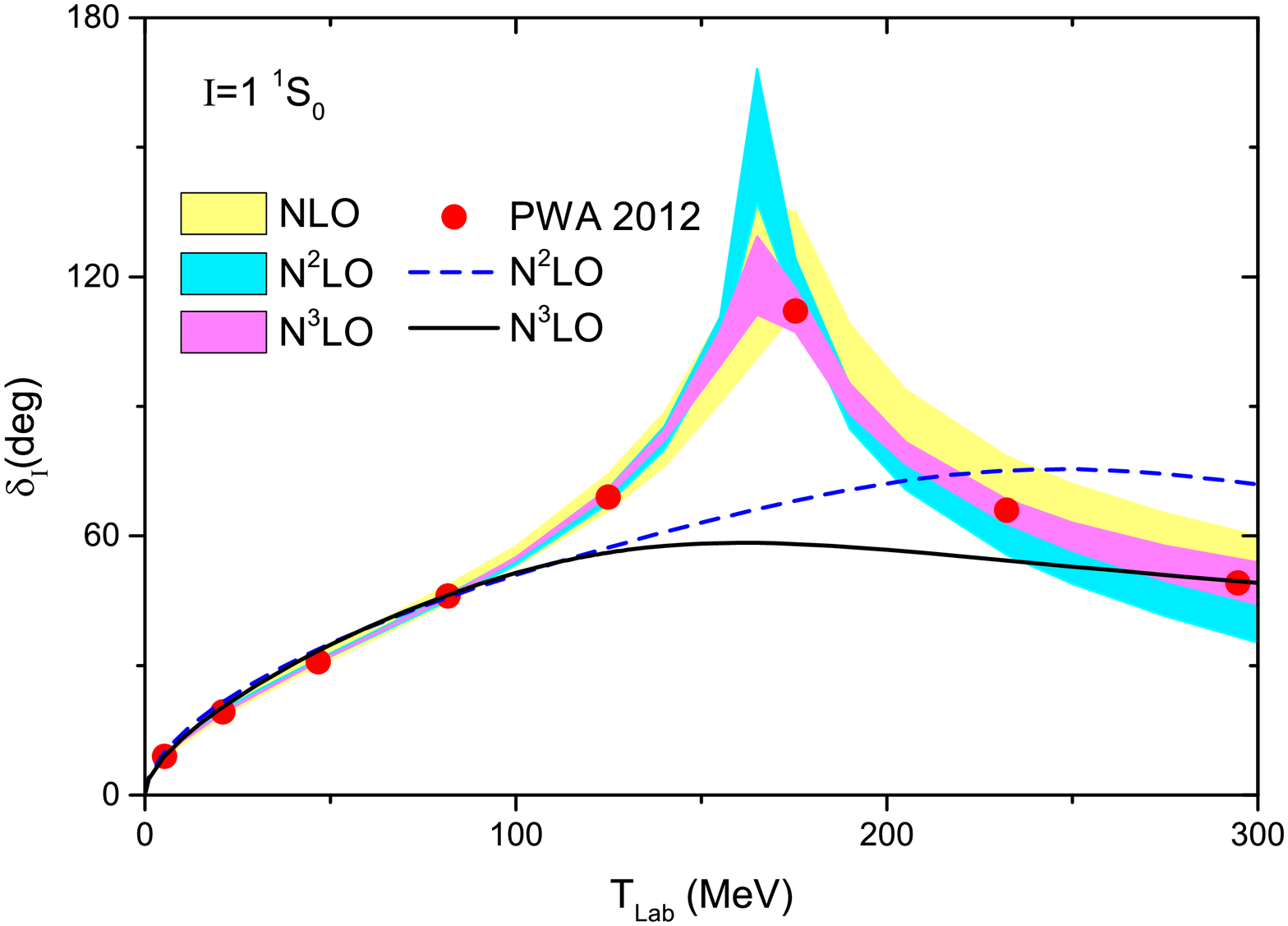}
\caption{Real and imaginary parts of the $^1S_0$ phase shift in the isospin $I=1$ channel.
The bands represent the fits to the PWA \cite{Zhou} (circles) at NLO, N$^2$LO, and N$^3$LO
from Ref.~\cite{Dai:2017}. The dashed and solid lines are refits at N$^2$LO and
N$^3$LO, respectively, utilized in the present work.
}
\label{fig:ppbar}
\end{center}
\end{figure}

\section{The reaction $\jpsi \to \gamma\ppbar$}
\label{sec:ppbar}

Due to the unusually large enhancement observed in the near-threshold $\ppbar$ invariant mass 
spectrum in the reaction $\jpsi \to \gamma\ppbar$ \cite{Bai:2003,Ablikim:2012,CLEOpsip},
it has been the topic of many studies and a variety of explanations for the strongly peaked 
spectrum have been suggested \cite{BesRev,Xu:2017}. In scenarios like ours, were FSI effects 
in the $\NNbar$ channel are assumed to be responsible for the enhancement, one faces a challenging 
task. There are measurements for several other decay channels where the produced 
$\NNbar$ state must be in the very same partial wave, the $^1S_0$, at least near treshold,
and accordingly, in principle, the same FSI effects should arise. 
This concerns the reactions $\jpsi \to \omega\ppbar$ \cite{Ablikim:ome}
and $\jpsi \to \phi\ppbar$ \cite{Ablikim:phi}, and also
$\psi (2S) \to \gamma\ppbar$ \cite{Ablikim:2012}. In none of these,
enhancements of a comparable magnitude were observed in the experiments.
So far, a few suggestions for a way out of this dilemma have been made
\cite{Kang:2015,Milstein:2017,Dedonder:2018}.
In our own work the emphasis was always on the isospin dependence.
Already in our initial studies \cite{Sibirtsev:2004,Haidenbauer:2006},
still based on the Migdal-Watson approximation and on the J\"ulich meson-exchange
$\NNbar$ potential \cite{Hippchen:1991,Mull:1991}, it was the isospin $I=1$ amplitude
that produced the large enhancement.
Then there is no conflict with the rather moderate enhancements observed in
the $\jpsi\to\omega\ppbar$  and $\jpsi\to\phi\ppbar$ channels, because in
those cases the produced $\ppbar$ system has to be in $I=0$ (assuming that
isospin is conserved in this purely hadronic decay).

\begin{figure}
\begin{center}
\includegraphics[height=55mm,clip]{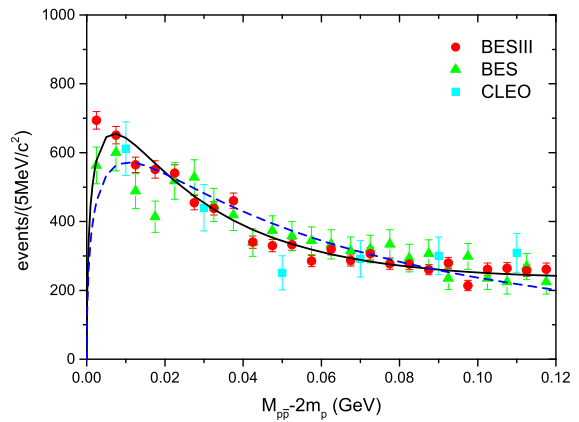}
\caption{$J/\psi \to \gamma \bar pp$ results with refitted $I=1$ $^1S_0$ amplitude, analoguous
to Ref.~\cite{Kang:2015}. Data are from Ref.~\cite{Ablikim:2012} (BESIII),
\cite{Bai:2003} (BES), and \cite{CLEOpsip} (CLEO). Note that the latter two are scaled to
those by the BESIII Collaboration by eye. 
}
\label{fig:nnbar}
\end{center}
\end{figure}

Indeed, in the decays $\jpsi \to \gamma\ppbar$ and $\psi (2S) \to \gamma\ppbar$
isospin is not conserved and, therefore, in principle, one can have any combination
of the $I=0$ and $I=1$ amplitudes. This freedom was exploited in a recent and more
refined study of $\jpsi$ decays by our group~\cite{Kang:2015}. In that work we not only
treated the FSI effects within a DWBA approach, but we also employed an $\NNbar$
potential that was derived within the framework of chiral effective field theory
up to N$^2$LO \cite{NNLO}.
Utilizing the ``standard'' hadronic combination for the $\ppbar$ amplitude, namely 
$T = T_{\ppbar} = (T^{I=0} + T^{I=1})/2$, for $\jpsi$ decay and
one with a predominant $I=0$ component, $T = (0.9\,T^{0} + 0.1\,T^{1})$
for $\psi (2S)$ decay allowed us to achieve a consistent description of the
$\gamma\ppbar$ spectrum for both decays~\cite{Kang:2015}.

Nonetheless, it should be said that we had to depart slightly from the $I=1$ $^{1}S_0$
$\NNbar$ amplitude as determined in the PWA of Zhou and Timmermans \cite{Zhou}.
However, already a rather modest modification of the interaction in the $I=1$ channel
-- subject to the constraint that the corresponding partial-wave cross sections
for $\ppbar \to \ppbar$ and $\ppbar \to \nnbar$ remain practically unchanged at low energies --
allowed us to reproduce the events distribution of the radiative $\jpsi$ decay,
and consistently all other decays \cite{Kang:2015}.

In the present work we repeat this exercise, employing now the new $\NNbar$ interaction \cite{Dai:2017}.
First of all, we want to see whether the same scenario holds for the improved
$\NNbar$ potential that is based on a different regularization scheme and that is
now calculated up to N$^3$LO. In addition we have to establish the $\jpsi \to \gamma\ppbar$
amplitude in the $I=0$ channel that enters into the calculation of the 2-step process,
see Eq.~(\ref{eq:LS}).
Results for the $\NNbar$ sector, i.e. the $I=1$ $^1S_0$ amplitude, are shown in Fig.~\ref{fig:ppbar}.
The parameters of the fit are summarized in Table~\ref{tab:para}.
\begin{table}[h!]
\renewcommand{\arraystretch}{1.5}
{\footnotesize
\begin{center}
\begin{tabular}{|c||c|c|c|}
\hline
                                     & N$^2$LO      & N$^3$LO              \\
\hline\hline
$\tilde{C}_{^{31}S_0}$~(GeV$^{-2}$)  &  0.1935(14)  &   0.3155(15)        \\
$C_{^{31}S_0}$~(GeV$^{-4}$)          & $-$1.8160(52)  &  $-$3.5235(101)         \\
$D^1_{^{31}S_0}$~(GeV$^{-6}$)        &  -           &  $-$8.0840(627)         \\
$D^2_{^{31}S_0}$~(GeV$^{-6}$)        &  -           &  10.0000(286)        \\
$\tilde{C}^a_{^{31}S_0}$~(GeV$^{-1}$)&  0.1733(25)  &   0.0230(33)         \\
$C^a_{^{31}S_0}$~(GeV$^{-3}$)        & $-$4.1780(21)  &  $-$3.1759(100)         \\
\hline
\end{tabular}
\caption{\label{tab:para} Low-energy constants at N$^2$LO and N$^3$LO, for 
the $\NNbar$ interaction in the $I=1$ $^1S_0$ partial wave.
Note that all parameters are in units of $10^4$, see Ref.~\cite{Dai:2017} for details.
}
\end{center}
}
\renewcommand{\arraystretch}{1.0}
\end{table}
Corresponding results for the $\ppbar$ invariant mass spectrum of the reaction
$\jpsi \to \gamma\ppbar$ are displayed in Fig.~\ref{fig:nnbar}. It is reassuring
to see that the results are basically the same as those reported in Ref.~\cite{Kang:2015} 
for the chiral N$^2$LO interaction. The presented results are for the combination 
$T = (0.4\,T^{0} + 0.6\,T^{1})$ that yields the lowest $\chi^2$ value in the fit. 
Note, however, that those for weights of the isospin amplitudes differing by, 
say, $\pm 0.1$ are very similar, even on a quantitative level. 

Interestingly, the modified potential in Ref.~\cite{Kang:2015} generates a bound state in
the $I=1$ $^{1}S_0$ partial wave which was not the case for the original interaction presented in Ref.~\cite{NNLO}.
For example, for the cutoff combination $\{\Lambda,\,\tilde\Lambda\}=$ \{450\,\text{MeV},\,500\,\text{MeV}\}
the bound state is located at $E_B=(-36.9-{\rm i}\, 47.2)\,\text{MeV}$, where the
real part denotes the energy with respect to the $\NNbar$ threshold. As noted 
in \cite{Kang:2015}, this bound state is not very far away from the position of the 
$X(1835)$ resonance found by the BES Collaboration in the reaction $\jpsi\to\gamma\eppc$
\cite{Ablikim:2005,Ablikim:2010,Ablikim:2016}.
However, the bound state in \cite{Kang:2015} is in the $I=1$ channel {\it and not}
in $I=0$ as advocated in publications of the BES Collaboration \cite{Ablikim:2005}
and of other authors \cite{Dedonder:2009,Milstein:2017}. 
The refit of the new $\NNbar$ potential \cite{Dai:2017} employed in the present study 
leads likewise to a bound state in the $I=1$ $^{1}S_0$ partial wave. The binding energies are 
$E_B=(-50.8-{\rm i}\, 40.9)\,\text{MeV}$ for the chiral N$^3$LO interaction
$E_B=(-2.1-{\rm i}\, 94.0)\,\text{MeV}$ for the chiral N$^2$LO interaction.
The former value is close to that found in our earlier work \cite{Kang:2015},
while the latter differs drastically. 
Once again, this illustrates the warning remarks in Ref.~\cite{Kang:2015} that, in general, 
any data above the reaction threshold, like the $\ppbar$ invariant mass spectrum or even phase
shifts, do not allow to pin down the binding energy reliably. 

\section{The reaction $\jpsi \to \gamma\eppc$}
\label{sec:results}


As already mentioned in the Introduction, in studies of $e^+e^-$ annihilation
to multipion states structures were observed around the $\NNbar$
threshold for several channels, specifically in $e^+e^- \to 3(\pi^+\pi^-)$,
$e^+e^- \to 2(\pi^+\pi^-\pi^0)$, and $e^+e^- \to 2(\pi^+\pi^-)\pi^0)$ \cite{BABAR,BABAR2,CMD,CMD2}.
An analyis of those structures performed by us \cite{Haidenbauer:2015} and by others
\cite{Milstein:2015} suggested that they could be simply a result of a threshold
effect due to the opening of the $\bar NN$ channel.
In that work we could estimate the contribution of the two-step process $e^+e^-\to \bar NN \to$
multipions to the total reaction amplitude rather reliably because cross-section measurements
for all involved processes were available in the literature. Specifically,
the amplitude for $e^+e^-\to \bar NN$ could be constrained from near-threshold data on the
$e^+e^-\to \bar pp$ cross section and the one for $\bar NN \to 5\pi,\,6\pi$ could be fixed
from available experimental information on the corresponding annihilation ratios \cite{Klempt}.
It turned out that the resulting amplitude for $e^+e^-\to\NNbar\to$ multipions was large enough
to play a role for the considered $e^+e^-$ annihilation channels and that it is possible to
reproduce the data quantitatively near the $\bar NN$ threshold in most of the considered
reaction channels \cite{Haidenbauer:2015}.

In case of $\jpsi\to \gamma\eppc$ we are not in such an advantageous situation. While
cross sections (or branching ratios) are available for $\ppbar \to \eppc$, so far only
event rates have been published for $\jpsi\to \gamma\eppc$ itself and for
$\jpsi\to \gamma\ppbar$. Thus, a reliable assessment of the magnitude of the two-step process
$\jpsi\to \gamma\ppbar\to \gamma\eppc$ cannot be given at present. Nonetheless, in the
following we provide a rough order-of-magnitude estimate and plausibility arguments why we
believe that the $\NNbar$ intermediate step should play an important role here. The
main and most important support comes certainly from the $\gamma\eppc$ data itself,
where a clear structure is seen at the $\NNbar$ threshold in the latest high-statistics
measurement by the BESIII Collaboration \cite{Ablikim:2016}.
In addition a comparison of the event rates for $\jpsi \to \gamma\ppbar$ and
$\jpsi \to \gamma\eppc$ with the cross sections for $\ppbar \to \ppbar$ in the $^1S_0$
partial wave and for $\ppbar \to \eppc$ suggests that the two-step process in question 
should be of relevance. 

Let us discuss the latter issue in more detail. 
With the central value of the branching ratio, $BR(\ppbar\to\eppc)=0.626$\% \cite{Amsler:2004},
the resulting cross sections at $p_{\rm lab} = 106$~MeV/c is $2.23$~mb, based on the total
annihilation cross section given in Ref.~\cite{BRrest}.
Though the branching ratio is tiny, at first sight, one has to compare the resulting cross section
with the relevant quantity, namely the $\ppbar$ elastic cross section in the $^1S_0$
partial wave. The latter is around $20$~mb in our $\NNbar$ potential \cite{Dai:2017}, but also in
the PWA \cite{Zhou}. Thus, the annihilation cross section for $\ppbar\to\eppc$ is roughly
a factor 10 smaller than that for $\ppbar \to \ppbar$.

When comparing the event rates one has to consider that the number of $\jpsi$ decay events
used in the $\gamma\eppc$ analysis \cite{Ablikim:2016} is roughly a factor five larger than
that in the $\gamma\ppbar$ paper \cite{Ablikim:2012}. Moreover, the bin size is different.
Combining those two aspects suggests a roughly five times larger rate for $\gamma\ppbar$,
based on the data shown in Refs.~\cite{Ablikim:2012,Ablikim:2016}, which mostly  
compensates for the factor of 10 reduction estimated above.

In the actual calculation we fix the constant $\tilde C_{\epp}$ in the $\NNbar \to \epp$
transition potential (cf. Eq.~(\ref{eq:VNN})) from the corresponding annihilation cross
section discussed above. Since there is no experimental information on the energy dependence,
we set the constant $C_{\epp}$ to zero. For the amplitude $A_{\jpsi\to\gamma\ppbar}$ we
employ the one described in Sect.~III, with $\tilde C_{\jpsi \to \gamma\bar NN}$ 
fixed to the most recent BESIII data \cite{Ablikim:2016}.
However, we allow for some variations of the overall magnitude because, as said above, only
event rates are available in this case.
The value for $C_{\jpsi \to \gamma\bar NN}$ obtained in the fit turned out to be very 
small so that we simply set it to zero.

Finally, the constants in the quantity $A^0_{\jpsi\to\gamma\eppc}$ (cf. Eq.~(\ref{eq:A0e})) 
are adjusted to the event rate for $\jpsi \to \gamma\eppc$. 
This term has to account for all other contributions to ${\jpsi \to \gamma\eppc}$, besides the
one with an intermediate $\gamma\NNbar$ state. Thus, it can have a relative phase as
compared to the contribution from the $\NNbar$ loop, i.e. the corresponding $C$'s can be 
complex valued. However, it turns out that optimal results are already achieved for real 
values of $\tilde C_{\jpsi \to \gamma\epp}$ and $C_{\jpsi \to \gamma\epp}$. 
In the fit we consider data in the range $1800 \ \rm{MeV}  \le E \le 1950$~MeV, i.e. in 
a region that encompasses more or less symmetrically the $\NNbar$ threshold. \\

\begin{figure}[htbp]
\begin{center}
\includegraphics[height=55mm,clip]{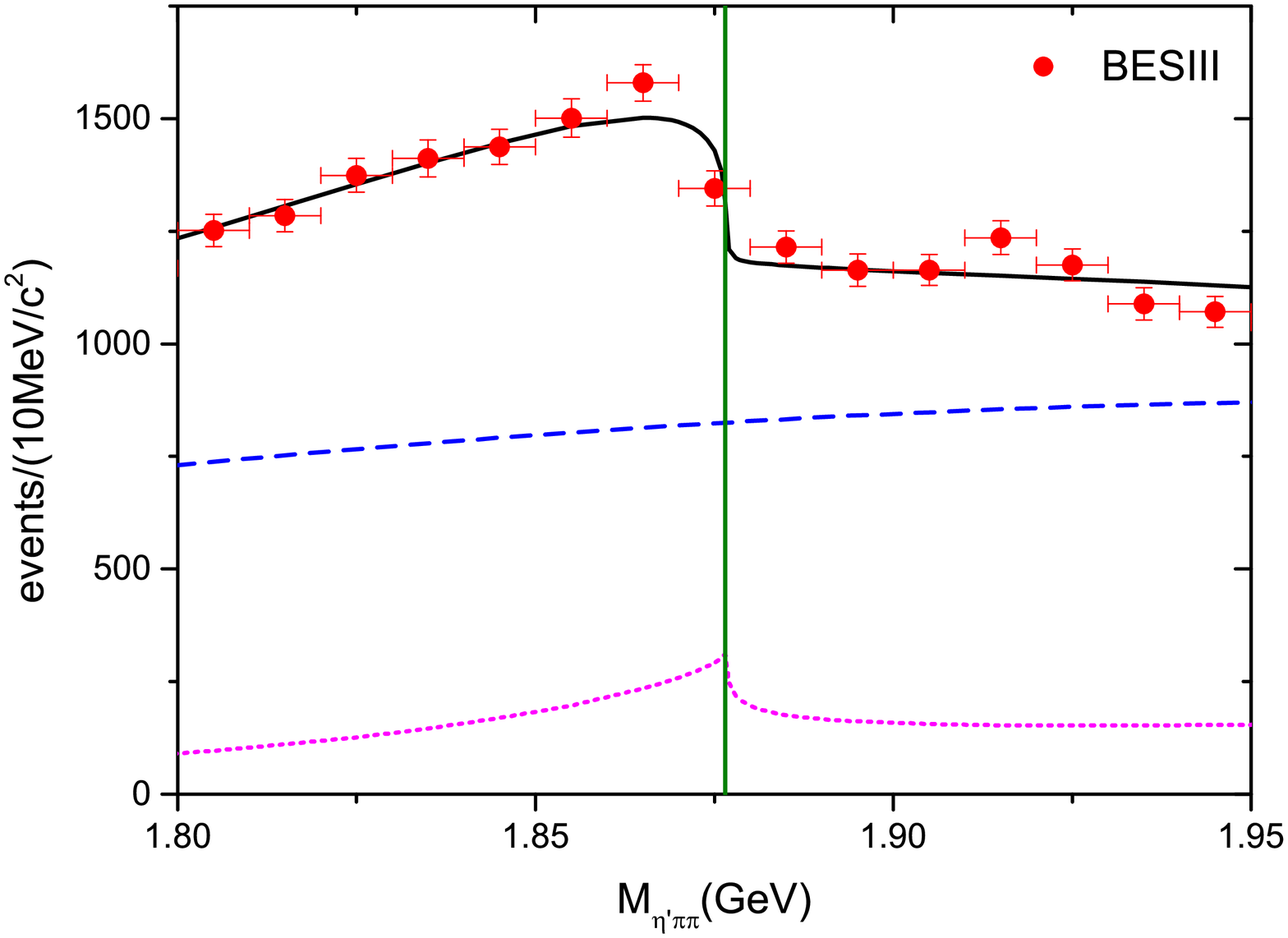}
\caption{The $\eppc$ invariant mass spectrum in the reaction $J/\psi \to \gamma \eta'\pi^+\pi^-$.
Results for the contribution from the $J/\psi \to \gamma \NNbar \to \gamma \eta'\pi^+\pi^-$ 
transition (dotted line) and the background term (dashed line) are shown, together with the 
full results (solid line). The N$^3$LO $\NNbar$ potential \cite{Dai:2017} is employed.
Data are from the BESIII Collaboration \cite{Ablikim:2016}.
The horizontal line indicates the $\ppbar$ threshold.
}
\label{fig:eppA}
\end{center}
\end{figure}

\begin{figure}[htbp]
\begin{center}
\includegraphics[height=55mm,clip]{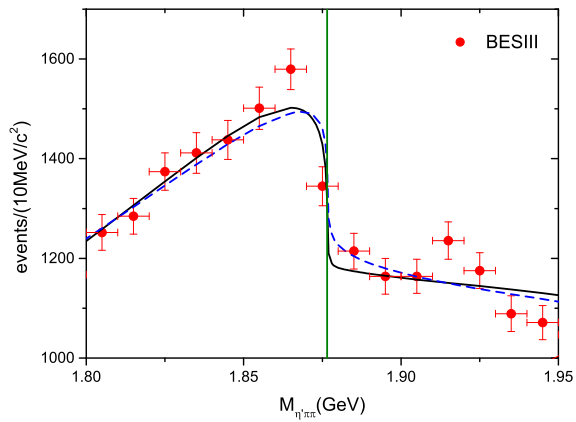}
\caption{Results for $J/\psi \to \gamma \eta'\pi^+\pi^-$ including background
term and $\bar NN \to \eta'\pi^+\pi^-$ transition amplitude for the
N$^2$LO (dashed line) and N$^3$LO (solid line) $\NNbar$ interactions. 
Data are from the BESIII Collaboration \cite{Ablikim:2016}.
The horizontal line indicates the $\ppbar$ threshold.
}
\label{fig:eppB}
\end{center}
\end{figure}


Our results for the reaction $\jpsi \to \gamma\eppc$ are presented in 
Figs.~\ref{fig:eppA} and \ref{fig:eppB}. They are based on the N$^2$LO and N$^3$LO 
EFT $\NNbar$ interactions with the cutoff $R=0.9$~fm ($\Lambda=438$~MeV), 
cf. Ref.~\cite{Dai:2017} for details.
Exploratory calculations for the other cutoffs considered
in Ref.~\cite{Dai:2017} turned out to be very similar. Like for $\NNbar$
scattering itself, much of the cutoff dependence is absorbed by the contact terms
($\tilde C_\nu$ and $C_\nu$ in Eqs.~(\ref{eq:VNN}) and (\ref{eq:A0N})) that are fitted
to the data so that the variation of the results for energies of, say, $\pm 50$~MeV
around the $\NNbar$ threshold is rather small. For consistency the momentum-space regulator
function as given in Eq.~(3.1) (right side) in Ref.~\cite{Dai:2017} is also attached 
to the transition potentials in Eqs.~(\ref{eq:VNN}) and (\ref{eq:A0N}), i.e.
to all quantities that depend on the $\NNbar$ momentum $q$. 

In Fig.~\ref{fig:eppA} the full results for the $\eppc$ invariant mass spectrum (solid line) 
are shown, together with the individual contributions from the $\jpsi \to \gamma\NNbar 
\to \gamma\epp$ transition (dotted line) and the background term (dashed line),
exemplary for our N$^3$LO interaction. By construction the background 
is a smooth function of the $\eppc$ invariant mass, whereas the contribution from the $\NNbar$ 
loop exhibits a pronounced cusp-like structure at the $\NNbar$ threshold. 
The (square of the) latter amplitude is roughly a factor 5 smaller. However, there is a 
sizable interference between the two amplitudes so that the coherent sum reflects the opening 
of (coupling to) the $\NNbar$ channel and leads to results for the invariant mass spectrum
that are very close to the measurements of the BESIII Collaboration. 

In Fig.~\ref{fig:eppB} we present the complete results for the N$^2$LO and N$^3$LO interactions,
on a scale similar to that in the BESIII publication \cite{Ablikim:2016}, 
cf. the inserts in Figs.~3 and 4 of that reference. First we note that the $\eppc$ invariant 
mass spectrum based on the two $\NNbar$ interactions is very similar around the $\NNbar$ 
threshold. It is also very similar to the fit within the {\it first model} considered in 
Ref.~\cite{Ablikim:2016} (cf. the corresponding Fig.~3). That model includes explicitly a 
$X(1835)$ resonance and simulates the effect of the $\NNbar$ channel via a Flatt\'e formula
\cite{Flatte}. Obviously, in our calculation the data can be described with the same
quality, but without such a $X(1835)$ resonance. The more elaborated treatment of the
coupling to the $\NNbar$ channel via Eq.~(\ref{eq:epp}) with the explicit inclusion 
of the $\NNbar$ interaction itself is already sufficient to generate 
an invariant-mass dependence in line with the data. 
 
For completeness, let us mention that a second resonance has been introduced 
in Ref.~\cite{Ablikim:2016} in the invariant-mass region covered by our study, namely 
an $X(1920)$, in order to reproduce a possible enhancement at the corresponding invariant 
mass suggested by two data points, cf. Fig.~\ref{fig:eppB}. 
Furthermore, a {\it second model} has been considered in Ref.~\cite{Ablikim:2016} where 
instead of the coupling to the $\NNbar$ channel an additional and rather narrow resonance 
was included, the $X(1870)$. In that scenario a slightly better description of the data
very close to the $\NNbar$ threshold could be achieved.  

Now the key question is, of course, are those structures seen in the experiment a signal 
for a $\NNbar$ bound state?  We did not find any near-threshold 
poles for our EFT $\NNbar$ interactions in the $^{1}S_0$ partial wave with $I=0$, i.e. 
the one relevant for the $\gamma\eppc$ channel, neither for the N$^2$LO potential presented 
in Ref.~\cite{NNLO} nor for the new N$^2$LO and N$^3$LO interactions \cite{Dai:2017} 
employed in the present calculation. 
As already discussed in the preceding section, there is only a pole in the $I=1$ case 
in the versions established in the study of the reactions $\jpsi \to \gamma\ppbar$.  

Thus, our results provide a clear indication that bound states are not necessarily required 
for achieving a quantitative reproduction of the observed structure in the $\eppc$ 
invariant-mass spectrum near the $\ppbar$ threshold. 
This is in contrast to other investigations in the literature. For example, bound states in 
the $I=0$ $^{1}S_0$ partial wave are present in the Paris $\NNbar$ potential \cite{Bennich} 
employed in Refs.~\cite{Dedonder:2009,Dedonder:2018} ($E_B = (-4.8-i\,26)$~MeV) and also in 
the $\NNbar$ interaction constructed in Ref.~\cite{Milstein:2017} ($E_B = (22-i\,33)$~MeV). 
In the latter case, the positive sign of the real part of $E_B$ indicates that the pole
found is actually located above the $\bar NN$ threshold (in the energy plane).
As discussed in Ref.~\cite{Milstein:2017}, the pole moves below the threshold when the imaginary 
part of the potential is switched off and that is the reason why it is referred to as bound state.

In this context, it is worth mentioning that no bound states or resonances were found 
in a study of the $\eta^\prime K\bar K$ system \cite{Martinez:2016} in
an attempt to explore in how far such states could be generated dynamically
as $\eta^\prime f_0(980)$- or $\eta^\prime a_0(980)$-like configurations.

Past studies suggest that there is a distinct difference in the amplitude for 
$\jpsi\to\gamma+${\it mesons}
due to the $\NNbar$ loop contribution in case of the absence/presence of a bound state.
Its modulus exhibits specific features, namely either a genuine cusp at the
$\NNbar$ threshold (cf. Fig.~\ref{fig:eppA}) or a rounded step and a maximum below the 
threshold. This was discussed in detail in Ref.~\cite{Haidenbauer:2015} in the context of 
the reaction $e^+e^- \to$ multipions (cf. Fig.~4 in that reference) and also in
Ref.~\cite{Milstein:2017}. However, in both studies the bound states in question belong to 
the special class discussed above, i.e. they are located above the $\bar NN$ threshold. 
 
\begin{figure}[htbp]
\begin{center}
\includegraphics[height=55mm,clip]{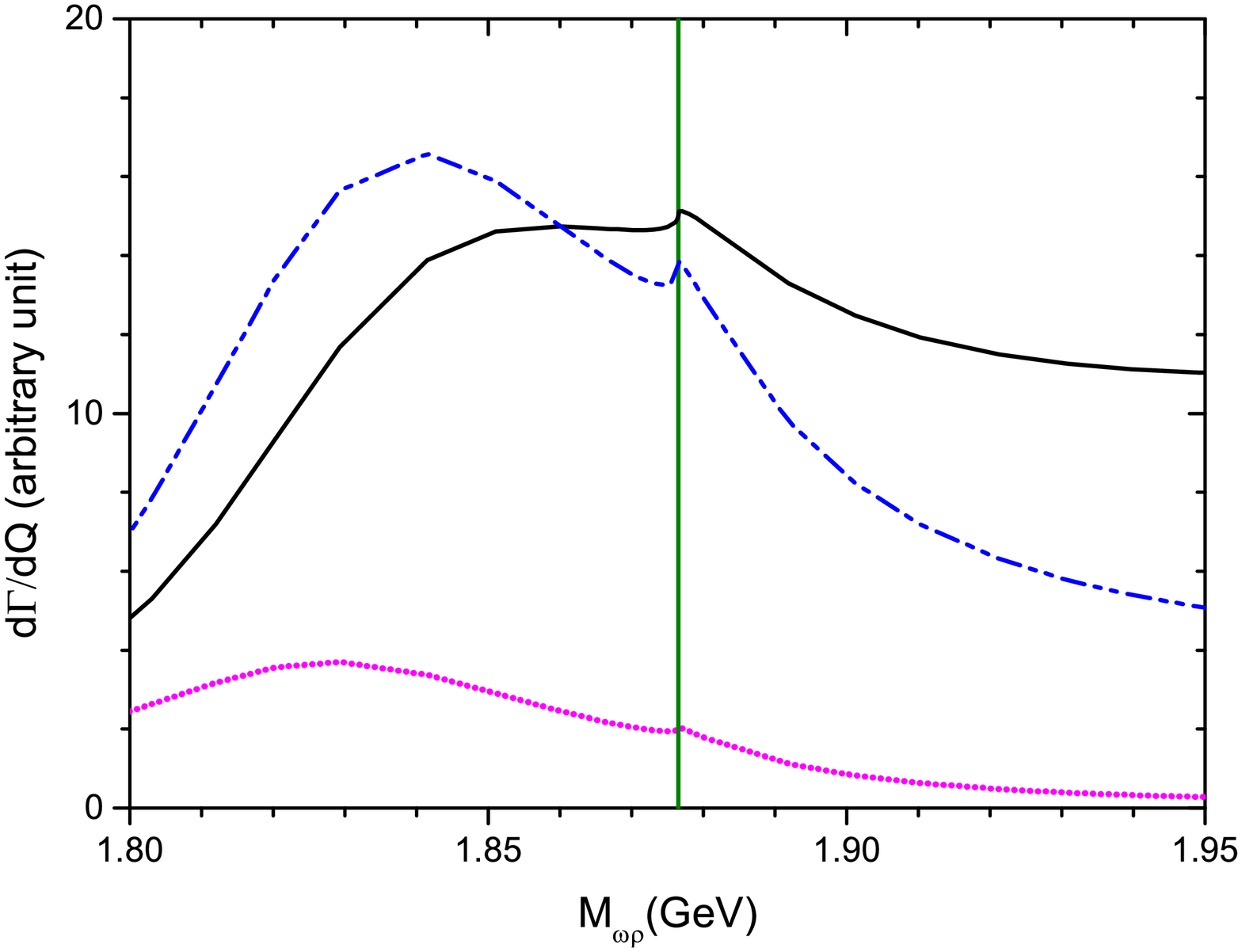}
\caption{Predicted $\omega\rho^0$ invariant-mass spectrum for $J/\psi \to \gamma \omega\rho^0$,
based on the N$^3$LO $\NNbar$ interaction described in Sect.~\ref{sec:ppbar}. 
The contribution from the $J/\psi \to \gamma \NNbar \to \gamma \omega\rho^0$ 
transition alone (dotted line) and with two arbitrary choices for the background 
term included (dashed and solid lines) are shown. 
The horizontal line indicates the $\ppbar$ threshold.
}
\label{fig:omr}
\end{center}
\end{figure}

In order to illustrate what happens for the case of a ``regular'' bound state we present 
here an exemplary calculation based on the $I=1$  $^1S_0$ partial wave of our N$^3$LO 
potential, where the binding energy is $(-50.8-{\rm i}\, 40.9)\,\text{MeV}$, cf. Sect.~\ref{sec:ppbar}. 
A $\jpsi$ decay reaction where the corresponding $\NNbar$ loop could contribute is,
for example, $\jpsi \to \gamma\omega\rho^0$. Pertinent predictions are shown in
Fig.~\ref{fig:omr}. Obviously, the invariant-mass dependence of the loop (dotted line)
is fairly different from the one of the $I=0$ amplitude, cf. dotted line in Fig.~\ref{fig:eppA}. 
Specifically, there is a clear enhancement in the spectrum around $50$~MeV below the $\NNbar$
threshold reflecting the presence of the $\NNbar$ bound state. Due to the fairly
large width ($\Gamma = -2\, {\rm Im} E_B$) the structure is not very pronounced. 
Of course, the final signal will be strongly influenced and modified by the interference 
with the background amplitude, as testified by the results presented above for the $\eppc$ case. 
For demonstrating this we include also results for two different but arbitrary choices
for the background term, see the dashed and solid lines in 
Fig.~\ref{fig:omr}. Of course, in case that the $\NNbar$ bound state is more narrow then
the signal will be certainly more pronounced. 
Note that the decay $\jpsi \to \gamma\omega\rho^0$ has been already measured by
the BES Collaboration \cite{Ablikim:2008}. However, the statistics is simply too low
for drawing any conclusions. It would be definitely interesting to revisit this 
reaction in a future experiment. 

\section{Conclusions}
\label{sec:summary}
We analyzed the origin of the structure associated with the $X(1835)$ resonance, observed
in the reaction $\jpsi \to\gamma\eppc$. Specific emphasis was put on the $\eppc$ invariant
mass spectrum around the $\bar pp$ threshold, where the most recent BESIII measurement
\cite{Ablikim:2016} provided strong evidence for an interplay of the $\eppc$ and $\ppbar$
channels.

Motivated by this experimental observation, we evaluated the contribution of the two-step
process $\jpsi\to \gamma\ppbar\to \gamma\eppc$ to the total reaction amplitude.
The amplitude for $\jpsi\to \gamma \ppbar$ was constrained from corresponding data
by the BESIII Collaboration, while for $\bar NN \to \epp$ we took available
branching ratios for $\ppbar \to \eppc$ as guideline.
Combining the contribution of this two-step process with a background amplitude, that
simulates other transition processes which do not involve an $\gamma\NNbar$ intermediate state,
allowed us to achieve a quantitative reproduction of the data near the $\ppbar$ threshold.
In particular, the structure detected in the experiment emerges as a threshold effect.
It results from an interference of the smooth background amplitude with the strongly
energy-dependent two-step contribution, which itself exhibits a cusp-like behavior at
the $\bar NN$ threshold.

The question whether there is an evidence for a $\NNbar$ bound state is discussed, but no
firm conclusion could be made. While in our own calculation such states are not present,
and are also not required for describing the data for the reaction
$\jpsi \to\gamma\eppc$, contrary claims have been brought forth in the 
literature \cite{Milstein:2017,Dedonder:2018}.
In any case, it should be said that the possibility that a genuine resonance is
ultimately responsible for the structure observed in the experiment cannot be
categorically excluded based on an analysis like ours. Yet,
our calculation provides a strong indication for the important role played by the $\NNbar$
channel in the $\jpsi\to \gamma\eppc$ decay for energies around its threshold
and we consider the fact that it yields a natural and quantitative description of
the structure observed in the invariant mass spectrum as rather convincing.

Data with improved resolution around the $\ppbar$ threshold could possibly help to 
shed further light on the relation of a possible $X(1835)$ with the $\ppbar$ channel. 
An absolute determination of the relevant
invariant-mass spectra would certainly put stronger constraints on the question
whether the intermediate $\ppbar$ state can play such an important role as suggested
by the present study.
In addition, we believe that an analogous measurement for channels like
$\jpsi\to \gamma\eta\pi^+\pi^-$ could be very instructive.
Indeed, this has been already recommended around the time when first evidence for the
$X(1835)$ was reported \cite{Zhu:2005}. The branching ratio for $\ppbar\to \eta\pi^+\pi^-$
is more than a factor two larger than for $\eppc$ \cite{Abele:1997} which would
enhance the role played by the $\ppbar$ channel. On the other hand, if the count
rates for $\jpsi\to \gamma\eta\pi^+\pi^-$ turn out to be much larger than those
for $\gamma\eppc$ \cite{Zhu:2005,PDG} then the effect from the transition to
$\ppbar$ should be strongly reduced or even disappear.

Finally, we want to mention that there are data on $\jpsi \to \omega\eta\pi^+\pi^-$ 
\cite{Ablikim:2011} and $\jpsi \to \phi\eta\pi^+\pi^-$ \cite{Ablikim:2014}. For the 
latter, $\eta\pi^+\pi^-$ invariant masses corresponding to the $\ppbar$ threshold 
are already close to boundary of the available phase space and, therefore, no
appreciable signal is expected. In case of $\jpsi \to \omega\eta\pi^+\pi^-$
the BES\-III Collaboration sees a resonance-like enhancement at
$1877.3\pm 6.3^{+3.4}_{-7.4}$~MeV \cite{Ablikim:2011} which coincides
almost perfectly with the $\ppbar$ threshold. However, the invariant-mass 
resolution of the present data is only $20$ MeV/c$^2$. Moreover, it is our
understanding that non-$\omega$ (background) events are not well separated in 
the data presented in Ref.~\cite{Ablikim:2011}. 
These two issues handicap a dedicated analysis for the time being. 
Clearly, new measurements with higher statistics could be indeed rather useful
for providing further information on the role that the (opening of the) $\NNbar$
channel plays for the reaction in question.

\vspace{\baselineskip}
\section*{Acknowledgements}
We acknowledge discussions with Dieter Grzonka on general aspects related to
the data analysis. 
This work is supported in part by the DFG and the NSFC through
funds provided to the Sino-German CRC 110 ``Symmetries and
the Emergence of Structure in QCD'' (DFG grant. no. TRR~110)
and the VolkswagenStiftung (grant no. 93562).
The work of UGM was supported in part by The Chinese Academy
of Sciences (CAS) President's International Fellowship Initiative (PIFI) (grant no.~2018DM0034).


\end{document}